\documentclass[12pt, a4paper]{article}
\usepackage{a4wide}
\usepackage{multirow}
\usepackage{amsmath}
\usepackage{amssymb}
\usepackage{amsfonts}
\usepackage{graphicx}
\date{\today}

\begin{document}
\title{9-variable Boolean Functions with Nonlinearity 242\\ in the Generalized Rotation Class\thanks{This work is based on the conference papers~\cite{kayu} and~\cite{kayu08}.}}

\author{
Sel\c{c}uk Kavut$\;^\dagger$ and Melek Diker Y{\"u}cel
\footnote{Department of Electrical Engineering and Institute of Applied Mathematics, Middle East Technical University -- ODT\"U, 
06531 Ankara, Turkey. Email: $\{$kavut, melekdy$\}$@metu.edu.tr .}
}
\date{}
\maketitle
\newcommand{\qed}{\hfill \rule{2mm}{2mm}}
\newcommand{\pf}{{\bf Proof: }}
\newtheorem{definition}{Definition}
\newtheorem{construction}{Construction}
\newtheorem{theorem}{Theorem}
\newtheorem{lemma}{Lemma}
\newtheorem{proposition}{Proposition}
\newtheorem{remark}{Remark}
\newtheorem{corollary}{Corollary}
\newtheorem{example}{Example}
\newtheorem{algorithm}{Algorithm}

\begin{abstract}
\noindent In 2006, 9-variable Boolean functions having nonlinearity 241, which is strictly greater than the bent concatenation bound of 240, have been discovered in the class of Rotation Symmetric Boolean Functions (RSBFs) by Kavut, Maitra and Y{\"u}cel.
To improve this nonlinearity result, we have firstly defined some subsets of the $n$-variable Boolean functions as the ``generalized classes of $k$-RSBFs and $k$-DSBFs ($k$-Dihedral Symmetric Boolean Functions)", where $k$ is a positive integer dividing $n$ and $k$-RSBFs is a subset of $l$-RSBFs if $k<l$. Secondly, utilizing the steepest-descent like iterative heuristic search algorithm used previously to identify the 9-variable RSBFs with nonlinearity 241, we have made a search within the classes of 3-RSBFs and 3-DSBFs. The search has accomplished to find 9-variable Boolean functions with nonlinearity 242 in both of these classes. It should be emphasized that although the class of 3-RSBFs contains functions with nonlinearity 242; 1-RSBFs or simply RSBFs, which is a subset of 3-RSBFs, does not contain any. This result also shows that the covering radius of the first order Reed-Muller code \textit{R}(1, 9) is at least equal to 242. Thirdly, motivated by the fact that RSBFs are invariant under a special permutation of the input vector, we have classified all possible permutations up to the linear equivalence of Boolean functions that are invariant under those permutations. Specifically, for 9-variable Boolean functions, $9!$ possible permutations are classified into 30 classes; and the search algorithm identifies some of these classes as \textit{rich}. The rich classes yield new Boolean functions with nonlinearity 242 having different autocorrelation spectra from those of the functions found in the generalized 3-RSBF and 3-DSBF classes. However, there is no zero value in the Walsh spectra of these functions; hence, none of them can be linearly transformed to 9-variable balanced functions with nonlinearity 242.\\ 


\end{abstract}
\date{}

\newpage
\section{Introduction}
\noindent On odd number of input variables $n$, constructing Boolean functions with maximum possible nonlinearity is an unsettled open problem in the area of cryptography and combinatorics. The problem is also related to the upper bound $\lfloor 2^{n-1}-2^{\frac{n}{2}-1} \rfloor$ on the covering radius of the first order Reed-Muller code~\cite{TOR78}, which is later improved~\cite{HO97} as $2\lfloor 2^{n-2}-2^{\frac{n}{2}-2}\rfloor$. Boolean functions on even number of input variables $n$, attaining maximum nonlinearity of ($2^{n-1}-2^{\frac{n}{2}-1}$) are called the bent functions~\cite{RO76}. For odd $n$, the nonlinearity value ($2^{n-1}-2^{\frac{n-1}{2}}$) is known as the \textit{bent concatenation bound}, since the concatenation of two bent functions on $(n-1)$ variables yields $n$-variable functions achieving this bound. In Table 1, we present the bent concatenation bound for $7 \leq n \leq 15$, together with recent nonlinearity results.

\vspace{1.8 mm}
\begin{center}
\noindent \small{\textbf{Table 1}. Summary of Nonlinearity Results for $n = 7, 9, 11, 13, 15$.} 
\begin{tabular}{| c | c | c | c | c | c |}
  \hline
  \small{$n$}    & \small{7} & \small{9} & \small{11}    & \small{13}     & \small{15}        \\ \hline
  Bent Concatenation Bound: $2^{n-1}-2^\frac{n-1}{2}$ & 56 & 240  & 992  & 4032 & 16256 \\ \hline
  Nonlinearity Results in~\cite{kavut}                  & $-$ & 241	 & 994  & 4036 & 16264 \\ \hline
	Our Nonlinearity Results 															& $-$ & 242  & 996  &	4040 & 16272 \\ \hline
	Patterson-Wiedemann Construction~\cite{PW83} 					& $-$ & $-$ & $-$ &	$-$ & 16276 \\ \hline
	Upper Bound~\cite{HO97} 															& 56 & 244  & 1000 &	4050 & 16292 \\ \hline
  
\end{tabular}	
\end{center}	
\vspace{2 mm}

\normalsize{}For odd $n \leq  7$, it is known that the maximum nonlinearity is equal to the bent concatenation bound~\cite{ER72, JJ80}. Clearly, as the number of $n$-variable Boolean functions ($2^{2^n}$) increases super-exponentially as $n$ increases, exhaustive search of the whole space is not feasible for $n \geq 7$ with currently available hardware. Therefore, for any search attempt, different subclasses of Boolean functions are always significant and interesting.

In 1983, Patterson and Wiedemann~\cite{PW83} demonstrated a construction in the idempotent class, of 15-variable Boolean functions with nonlinearity 16276 (exceeding the bent concatenation bound by 20), using combinatorial techniques and search methods. Since then, it has been possible to get functions with nonlinearity ($2^{n-1}-2^\frac{n-1}{2}+20\times 2^\frac{n-15}{2}$) for odd $n \geq 15$, which exceeds the bent concatenation bound by $20\times 2^\frac{n-15}{2}$. Until 2006, the maximum nonlinearity known for the cases of $n = 9, 11, 13$ was equal to the bent concatenation bound. In 2006, 9-variable Rotation Symmetric Boolean Functions (RSBFs) with nonlinearity 241 ($=2^{9-1}-2^\frac{9-1}{2}+1$) were discovered~\cite{kavut}, which led to the construction of functions with nonlinearity exceeding the bent concatenation bound by $1\times 2^\frac{n-9}{2}$, for odd $n \geq 9$ (see Table 1). Such functions were attained utilizing the steepest-descent like iterative algorithm that first appeared in~\cite{KM2} and then suitably modified in~\cite{kavut} for a search in the class of RSBFs. 

The class of Rotation Symmetric Boolean Functions seems to be \textit{rich} in terms of highly nonlinear functions and there is a close relation between RSBFs and idempotents~\cite{FE98, FE99}. Considering a Boolean function $f$ as a mapping from $GF(2^n)\rightarrow GF(2)$, the functions for which $f(\alpha^2) = f(\alpha)$ for any $\alpha \in GF(2^n)$, are referred to as idempotents. As pointed out in~\cite{FE98, FE99}, the idempotents can be regarded as Rotation Symmetric Boolean Functions with proper choice of basis. In~\cite{PW83}, 15-variable Patterson-Wiedemann functions having nonlinearity 16276 are also identified in the idempotent class. 

As the space of the RSBF class is much smaller ($\approx 2^{\frac{2^n}{n}}$) than the total space of Boolean functions ($2^{2^n}$) on $n$ variables, it is possible to exhaustively search the space of RSBFs up to a certain value of $n$. In~\cite{KMSY}, an exhaustive search carried out for the whole space ($2^{60}$) of 9-variable RSBFs exploiting some combinatorial results related to the Walsh spectra of RSBFs, has shown that there is no RSBF having nonlinearity greater than 241. Consequently, in order to find functions with higher nonlinearity, one needs to increase the search space.

Motivated by this fact, we have firstly proposed the generalized $k$-RSBFs, as functions which satisfy $f(\alpha^{2^k}) = f(\alpha)$, where $1 \leq k \; |\; n$. Note that if $k=1$, the resulting functions are the same as idempotents; whereas for $k=n$ the entire space of $n$-variable Boolean functions is covered. In the space of generalized $k$-RSBFs, imposing the condition of invariance under the action of dihedral group, we have defined the class of generalized $k$-DSBFs as a subset of $k$-RSBFs.

Secondly, we have used the steepest-descent like iterative algorithm in~\cite{kavut} for a search in the generalized 3-DSBF and 3-RSBF classes. This search has successfully ended up with 9-variable functions in both of these classes, having nonlinearity 242, absolute indicator values 32, 40 and 56. This result shows that the covering radius of the first order Reed-Muller code \textit{R}(1, 9) is at least 242. This result is also important for $n=11$ and $n=13$, since the bent concatenation of 9-variable functions with nonlinearity 242 leads to the construction of 11-variable and 13-variable functions with nonlinearity ($2^{n-1}-2^\frac{n-1}{2}+2\times 2^\frac{n-9}{2}$), which exceeds the bent concatenation bound by $2\times 2^\frac{n-9}{2}$ (see Table 1). However, we should mention that for odd $n \geq  15$, the nonlinearity ($2^{n-1}- 2^\frac{n-1}{2} +20\times 2^\frac{n-15}{2}$) given in~\cite{PW83} that can be obtained by concatenating 15-variable Patterson-Wiedemann functions is still greater than the nonlinearity ($2^{n-1}-2^\frac{n-1}{2}+2\times 2^\frac{n-9}{2}$).

Thirdly, knowing the fact that RSBFs, DSBFs, as well as the generalized $k$-RSBFs and $k$-DSBFs are invariant under some special types of permutations on input vectors, we have focused on the same search problem from a different direction and considered the possibility of other `\textit{rich}' classes that are invariant under some permutations. Linearly equivalent Boolean functions have the same nonlinearity; therefore, while searching for highly nonlinear functions, it is quite logical to classify all $n!$ permutations up to the linear equivalence of Boolean functions that are invariant under them. More specifically, for 9-variable Boolean functions, we have classified $9!$ many permutations into 30 classes which are different up to the linear equivalence of Boolean functions that are invariant under them. Then for each class, by picking up a representative permutation arbitrarily, we have searched the corresponding set of Boolean functions. Consequently, in some of these sets, we have obtained 9-variable Boolean functions with nonlinearity 242 and absolute indicator values 40, 48 $\&$ 56. So, our aim of defining other `\textit{rich}' classes is accomplished. We note that, however, the functions presented in this paper do not contain any zero in their Walsh spectra, and hence, they cannot be linearly transformed to balanced functions. 

In the following section, after reviewing some basic definitions related to Boolean functions, we present preliminaries of permutation group actions. In Section 3, we introduce the generalized rotation symmetric and dihedral symmetric Boolean functions. Classification of permutations on inputs of 9-variable Boolean functions, with respect to the linear equivalence of Boolean functions that are invariant under them, is presented in Section 4. Different results related to 9-variable Boolean functions with nonlinearity 242 are presented in both Section 3 and Section 4. Finally, some additional 11 and 13-variable DSBFs, which are attained by the steepest-descent like search algorithm with nonlinearities 994 and 4036 respectively, are presented in Section 3. It should be noticed that those functions have exactly the same nonlinearities, as those would be obtained by  concatenating 9-variable functions with nonlinearity 241.

\section{Preliminaries}

\subsection{Boolean Functions}

\noindent An $n$-variable Boolean function $f(x)$ produces a single-bit result for each $n$-bit input vector $x = (x_0, \ldots, x_{n-1})$, which may be considered as a mapping from $\{0, 1\}^n$ into $\{0, 1\}$. $f(x)$ is basically represented by its \textit{truth table}, that is, a binary vector of length $2^n$, $$f = [f (0,0,\ldots,0), f (1,0,\ldots,0), f (0,1,\ldots,0), \ldots, f (1,1,\ldots,1)].$$ We represent the set of all $n$-variable Boolean functions by $\mathcal{B}_n$; clearly $|\mathcal{B}_n| = 2^{2^n}$. A binary vector $g$ has the \textit{Hamming weight} $wt(g)$ equal to the number of its nonzero elements. The \textit{Hamming distance} between two binary vectors $g$ and $h$, both having the same length, is defined as the number of places for which $g$ and $h$ differ, i.e., $d(g, h) = wt(g\oplus h)$, where $\oplus$ denotes the addition over $GF(2)$. An $n$-variable Boolean function $f$ is called \textit{balanced} if $wt( f ) = 2^{n-1}$. 

The \textit{algebraic normal form} (ANF) of a Boolean function is defined as its unique representation in the form of a multivariate polynomial over $GF(2)$, $$f(x_0, \ldots, x_{n-1}) = c \oplus \bigoplus_{0 \leq i \leq n-1} a_i x_i \oplus \bigoplus_{0 \leq i < j \leq n-1} a_{ij} x_i x_j \oplus \ldots \oplus a_{01\ldots n-1} x_0 x_1 \ldots x_{n-1},$$ where the coefficients $c, a_i, a_{ij}, \ldots, a_{01\ldots n-1} \in  \{0, 1\}$. The \textit{algebraic degree}, or simply the degree of $f$, is the number of variables in the highest order product term with nonzero coefficient, which is denoted by $deg( f )$.

A Boolean function is called \textit{affine} if its degree is at most one. The set of all $n$-variable affine functions is represented by $A_n$. The nonlinearity of an $n$-variable Boolean function $f$ is defined as its minimum distance to any affine function, i.e., $$nl(f) = \min_{g \in A_n} (d(f, g)).$$ Boolean functions used in cryptographic systems must be highly nonlinear to resist Best Affine Approximation (BAA) and correlation attacks~\cite{CATR00, CD91}. 

The \textit{Walsh transform} of an $n$-variable Boolean function $f (x)$ is an integer valued function over $\{0, 1\}^n$ defined as $$W_f(w) = \sum_{x \in \{0,1\}^n} (-1)^{f(x)} (-1)^{<x,w>},$$ where $x = (x_0, \ldots, x_{n-1}), w = (w_0, \ldots, w_{n-1}) \in \{0, 1\}^n$ and $<x,w>= x_0w_0 \oplus \ldots \oplus x_{n-1}w_{n-1}$. 

The nonlinearity of an $n$-variable Boolean function $f(x)$ can be alternatively expressed by means of its Walsh spectrum, i.e., $$nl(f)=\frac{1}{2}(2^n - \max_{w \in \{0,1\}^n}|W_f(w)|.$$

The autocorrelation function of $f(x)$ is given by $$r_f(d) = \sum_{x \in \{0,1\}^n} (-1)^{f(x)} (-1)^{f(x \oplus d)},$$ where $d=(d_0, \ldots, d_{n-1}) \in \{0, 1\}^n$. The autocorrelation value having maximum magnitude except the origin is also known as the absolute indicator~\cite{ZZ95} and denoted as $$\Delta_f = \max_{d\neq {(0, \ldots, 0)} \in \{0,1\}^n}|r_f(d)|.$$

A Boolean function is balanced if and only if its Walsh spectrum value is zero at the origin. On the other hand, if an unbalanced Boolean function $g(x)$ contains a zero in its Walsh spectrum except the origin, say $W_g(u)=0$ and $u \neq (0,\ldots, 0)$, it can be linearly transformed into a balanced function $f(x) = g(x) \oplus <x,u>$, which has the same nonlinearity and absolute indicator; i.e., $nl(f)=nl(g)$ and $\Delta_f = \Delta_g$. 

\subsection{Group Action by Permutation Groups}

A group $G$ is said to act on a set $X$ if there is a mapping $\phi : G\times X \rightarrow X$ denoted as $g \cdot x$, which satisfies the following two axioms for all elements $x \in  X$. 

\begin{enumerate}
	\item $e \cdot x = x$ where $e$ stands for the identity element of $G$.
	\item $g \cdot (h \cdot x) = (gh) \cdot x$ for all $g, h \in G$. 
\end{enumerate}

\noindent The mapping $\phi$ is called the \textit{group action} and the set $X$ is called a $G$-set. The orbit of $x$ is defined as the set $G(x) = \{g \cdot x\; |\; g \in   G\}$, i.e., the group action moves $x$ to its orbit. As the set of orbits of $X$ under the action of $G$, denoted by $\mathcal{G}$, constitutes a partition of $X$, the corresponding equivalence relation is defined by $x \sim y$ iff there exists a $g \in G$ such that $g \cdot x = y$. Hence, the orbits form the equivalence classes under this relation.

Let $G$ be a permutation group acting on $\{0, 1\}^n$, and consider the class of $n$-variable Boolean functions which are invariant under the action of $G$, i.e., any Boolean function $f$ in the class satisfies the condition for each $x \in  \{0, 1\}^n, f(x) = f(y)$, for all $y \in G(x)$. As a consequence of the invariance property, the class composes a subclass of $\mathcal{B}_n$, and knowing the number of orbits, i.e., $|\mathcal{G}|$, it contains $2^{|\mathcal{G}|}$ many $n$-variable Boolean functions, each satisfying the given condition. The value of $|\mathcal{G}|$ can be determined using Burnside's Lemma.

\begin{lemma}
[Burnside's Lemma] Let G be a group of permutations acting on a set X, and $fix_X (g) = \{x \in  X\; |\; g \cdot x = x\}$ for each $g \in G$. Then
the number of orbits induced on X is given by
$$\frac{1}{|G|}\sum_{g \in G}|fix_X(g)|.$$
\end{lemma}

Let us represent an orbit by its lexicographically first element denoted by $\Lambda_i$. The Boolean function $f$ which is invariant under the action of $G$, can be represented by $(f(\Lambda_0), \ldots, f (\Lambda_{|\mathcal{G}|-1}))$, where $\Lambda_0, \ldots, \Lambda_{|\mathcal{G}|-1}$ are again arranged lexicographically. Clearly, this representation is shorter than the truth table of $f$. Further, it can be shown~\cite{SMC04} that $W_f(u) = W_f(v)$ if $u \in G(v)$, implying that the Walsh spectrum of $f$ can be at most $|\mathcal{G}|$ valued. Then, defining a $|\mathcal{G}|\times |\mathcal{G}|$ matrix $\mathcal{M}$ as $\mathcal{M}_{i,j} = \sum_{x \in G_{\Lambda_i}} {(-1)^{<x, \Lambda_j>}}$, the Walsh spectrum of $f$ can be calculated as~\cite{SMC04} $$W_f(\Lambda_j) = \sum_{i=0}^{|\mathcal{G}|-1}
(-1)^{f(\Lambda_i)} \, \mathcal{M}_{i, j}.$$

\section{Generalized Rotation and Dihedral Symmetric\\ Boolean Functions}

\noindent After briefly summarizing RSBFs, we propose the generalized classes of $k$-RSBFs and $k$-DSBFs in Definition 1 and Definition 2 respectively. Letting
$(x_0, x_1, \ldots,$ $x_{n-1}) \in$ $\{0,1\}^n$, the (left) $i$-cyclic shift operator ${\rho^i}_n$ on $n$-tuples is defined as $${\rho^i}_n(x_0, x_1, \ldots, x_{n-1}) = (x_{(0+i)\, \mathrm{mod}\, n}, \ldots, x_{(n-1+i)\, \mathrm{mod}\, n}),$$ for $1 \leq i \leq n$.

A Boolean function $f$ is called \textit{rotation symmetric} if for each
input $(x_0, \ldots,$ $x_{n-1}) \in \{0, 1\}^n, f({\rho^1}_n(x_0, \ldots, x_{n-1})) = f(x_0, \ldots, x_{n-1})$. That is, RSBFs are invariant under all cyclic rotations of the inputs. The
inputs of a rotation symmetric Boolean function can be divided into \emph{orbits} so
that each orbit consists of all cyclic shifts of one input. An orbit generated by
$(x_0, x_1, \ldots, x_{n-1})$ is denoted by $G_n(x_0, x_1, \ldots, x_{n-1}) = \{{\rho^i}_n(x_0, x_1, \ldots, x_{n-1})\;|$ $\;1\leq i\leq n\}$ and the number of such orbits is represented by $g_n$ ($\approx 2^{\frac{2^n}{n}}$). More specifically, $g_n$ is equal
to $\frac{1}{n}\sum_{t|n} \phi(t)2^\frac{n}{t}$~\cite{SM02}, where $\phi(t)$ is the Euler's phi-function. The total
number of $n$-variable RSBFs is $2^{g_n}$.

In the following, we define the generalized RSBFs as $k$-rotation symmetric
Boolean functions ($k$-RSBFs).

\begin{definition}
Let $1 \leq k \leq n, k\, |\, n$. An $n$-variable
Boolean function $f$ is called $k$-rotation symmetric if for each input $(x_0, \ldots, x_{n-1})$ $\in \{0, 1\}^n$, $f({\rho^k}_n(x_0, \ldots,$ $x_{n-1})) = f(x_0, \ldots, x_{n-1})$.
\end{definition}

\noindent As can be seen, the $k$-rotation symmetric Boolean functions are invariant
under $k$-cyclic rotations of inputs. Therefore, an orbit of a $k$-RSBF generated
by $(x_0, x_1, \ldots, x_{n-1})$ is ${G^k}_n(x_0, x_1, \ldots, x_{n-1}) = \{{\rho^i}_n(x_0, x_1, \ldots, x_{n-1})\tiny{\mbox{ }}|\tiny{\mbox{ }}i = k, 2k,$ $3k, \ldots, n\}$. For example, ${G^3}_9(001$ $001\; 111) = \{(001\; 001\; 111), (001\; 111\; 001),$ $(111\; 001\; 001)\}$.

If $g_{n,k}$ is the number of distinct orbits in the class of $k$-RSBFs of $n$ variables,
one can show that $g_{n,k} = \frac{k}{n} \sum_{t|\frac{n}{k}} \phi(t)2^{\frac{n}{t}} (\approx 2^{k\times \frac{2^n}{n}})$, where $\phi(t)$ is the Euler's phi
function.

The class of Dihedral Symmetric Boolean Functions (DSBFs)~\cite{MSD07}, a subset of the RSBF class, is invariant under the action of the dihedral group denoted by $D_n$. In addition to the (left) $i$-cyclic shift operator ${\rho^i}_n$ on $n$-tuples, which is defined previously, the dihedral group $D_n$ also includes the reflection operator $\tau_n(x_0, x_1, \ldots, x_{n-1}) = (x_{n-1}, \ldots, x_1, x_0)$. The $2n$ permutations of $D_n$ are then defined as $\{{\rho^1}_n, {\rho^2}_n, \ldots,$ ${\rho^{n-1}}_n, {\rho^n}_n, {\tau_n}{\rho^1}_n, {\tau_n}{\rho^2}_n, \ldots, {\tau_n}{\rho^{n-1}}_n,$ ${\tau_n}{\rho^n}_n\}$. The dihedral group $D_n$ generates
equivalence classes in the set $\{0,1\}^n$~\cite{FR}. Let $d_n$ be the number of such partitions.
The following proposition gives the exact count of $d_n$~\cite[page 184]{Harary},~\cite{MSD07}.

\begin{proposition}
Let $d_n$ be the total number of orbits induced by the dihedral
group $D_n$ acting on $\{0,1\}^n$. Then $d_n = g_n/2 + l$, where, $g_n = \frac{1}{n}\sum_{t|n} \phi(t)2^\frac{n}{t}$ is the
number of rotation symmetric classes\emph{~\cite{SM02}}, $\phi(t)$ is the Euler's phi-function and \\

\noindent $
l = \left\{
      \begin{array}{ll}
        \frac{3}{4}2^{\frac{n}{2}}, & \hbox{if $n$ is even,} \\
        2^{\frac{n-1}{2}}, & \hbox{if $n$ is odd.}
      \end{array}
    \right.
$
\end{proposition}

\noindent Since there are $2^{d_n}$ many $n$-variable DSBFs and $d_n \approx 2^\frac{2^n}{2n}$, a reduction in the size of the
search space over the size of RSBFs is provided.

\begin{definition}
Let $1 \leq k \leq n, k\, |\, n$. An $n$-variable
Boolean function $f$ is called $k$-dihedral symmetric if $f$ is invariant under the
group action ${D^k}_n = \{{\rho^{i}}_n, {{\tau}_n}{\rho^{i}}_n\tiny{\mbox{ }}|$ $i = k, 2k, 3k, ..., n\}$.
\end{definition}

\noindent As the class of DSBFs is a subspace of $k$-DSBFs, we call $k$-DSBFs
``generalized dihedral symmetric Boolean functions". One should observe that
$k$-DSBFs is a subspace of $k$-RSBFs.

When Proposition 1 is applied to $k$-dihedral symmetric functions, we obtain
the following corollary.

\begin{corollary}
Let $d_{n,k}$ be the number of distinct orbits, in the class of $k$-DSBFs
of $n$ variables. Then, $d_{n,k} = g_{n,k}/2 + l$, where, $g_{n,k} = \frac{k}{n} \sum_{t|\frac{n}{k}} \phi(t)2^{\frac{n}{t}}$ is the
number of $k$-rotation symmetric classes, $\phi(t)$ is the Euler's phi-function and\\

\noindent$
l = \left\{
  \begin{array}{ll}
    2^{\frac{n}{2}-1}, & \hbox{if $n$ is even, $k$ is even,} \\
    \frac{3}{4}2^{\frac{n}{2}}, & \hbox{if $n$ is even, $k$ is odd,} \\
    2^{\frac{n-1}{2}}, & \hbox{if $n$ is odd.}
  \end{array}
\right.
$
\end{corollary}

Table 2 compares the orbit counts of $k$-rotational classes, $k$-dihedral classes,
RSBFs and DSBFs for $k\, |\, n, n\le 15$.\\


\noindent \small{{\textbf{Table 2}. Comparison of the Orbit Counts $g_n, d_n, g_{n,k}$ and $d_{n,k}$ (for $n = 4, 6, \ldots, 15$, and all integers $k$, which divide $n$).}}
\begin{center}
\begin{tabular}{|c|c|c|c|c|c|c|c|c|}
  \hline

  \multicolumn{3}{|@{}r|}{$k$} & 2 & 3 & 4 & 5 & 6 & 7 \\
  \multicolumn{3}{|@{}l|}{\mbox{ $\; \, n$}} &   &   &   &   &   &   \\
  \hline

    \multirow{2}{*}{4} & $g_4=6$ & $g_{4,k}$ & 10 & -- & -- & -- & -- & -- \\
                       \cline{2-9}
                       & $d_4=6$ & $d_{4,k}$ & 7  & -- & -- & -- & -- & -- \\
                       \hline

    \multirow{2}{*}{6} & $g_6=14$ & $g_{6,k}$ & 24 & 36 & -- & -- & -- & -- \\
                       \cline{2-9}
                       & $d_6=13$ & $d_{6,k}$ & 16 & 24 & -- & -- & -- & -- \\
                       \hline

    \multirow{2}{*}{8} & $g_8=36$ & $g_{8,k}$ & 70 & -- & 136 & -- & -- & -- \\
                       \cline{2-9}
                       & $d_8=30$ & $d_{8,k}$ & 43 & -- & 76 & -- & -- & -- \\
                       \hline

    \multirow{2}{*}{9} & $g_9=60$ & $g_{9,k}$ & -- & 176 & -- & -- & -- & -- \\
                       \cline{2-9}
                       & $d_9=46$ & $d_{9,k}$ & -- & 104 & -- & -- & -- & -- \\
                       \hline

    \multirow{2}{*}{10} & $g_{10}=108$ & $g_{10,k}$ & 208 & -- & -- & 528 & -- & -- \\
                       \cline{2-9}
                        & $d_{10}=78$  & $d_{10,k}$ & 120 & -- & -- & 288 & -- & -- \\
                       \hline

    \multirow{2}{*}{12} & $g_{12}=352$  & $g_{12,k}$ & 700 & 1044 & 1376 & -- & 2080 & -- \\
                       \cline{2-9}
                        & $d_{12}=224$  & $d_{12,k}$ & 382 & 570  & 720  & -- & 1072 & -- \\
                       \hline

    \multirow{2}{*}{14} & $g_{14}=1182$ & $g_{14,k}$ & 2344 & -- & -- & -- & -- & 8256 \\
                       \cline{2-9}
                        & $d_{14}=687$  & $d_{14,k}$ & 1236 & -- & -- & -- & -- & 4224 \\
                       \hline

    \multirow{2}{*}{15} & $g_{15}=2192$ & $g_{15,k}$ & -- & 6560 & -- & 10944 & -- & -- \\
                       \cline{2-9}
                        & $d_{15}=1224$ & $d_{15,k}$ & -- & 3408 & -- & 5600  & -- & -- \\
                       \hline

\end{tabular}
\end{center}
\vspace{2 mm}

\normalsize As mentioned previously, we have utilized the steepest-descent like iterative search algorithm, which was used to discover 9-variable RSBFs having nonlinearity 241~\cite{kavut}. The details, which can be found in the related literature, are extraneous for present purposes. We note only that it is an efficient search technique with an outstanding ability to escape from local optima. 

\subsection{9-variable 3-DSBFs and 3-RSBFs}

We have applied the search to 9-variable 3-DSBFs for which the size of search space is $2^{104}$ (see Table 2). We have found several unbalanced
Boolean functions having nonlinearity 242. Among them there are two
different absolute indicator values, which are 32 and 40.
The following is the truth table of a 9-variable, 3-dihedral symmetric Boolean
function having nonlinearity 242, absolute indicator value 40, and algebraic
degree 7:

\begin{center}
\ttfamily{68B7EF2DA03B0D3EA00DB6A96DD99AEAFDB9C842B6D5DC8C4526CE0DD29020DB\\
B75FE3314568344E73688FF0CB2482E065231869E1AA4583765CC491F8A8DB12\\}
\end{center}

\noindent And, the function below is another 9-variable 3-DSBF having nonlinearity
242, absolute indicator value 32, and algebraic degree 7:

\begin{center}
\ttfamily{125425D30A398F36508C06817BEE122E250D973314F976AED58A3EA9120DA4FE\\
0E4D4575C42DD0426365EBA7FC5F45BE9B2F336981B5E1863618F49474F6FE00\\}
\end{center}

Using a computer system with Pentium IV 2.8 GHz processor and 256 MB
RAM, a typical run
of the search algorithm takes 1 minute and 17 seconds. We have carried out
100 runs, each each starting with a randomly chosen Boolean function in the space of 9-variable 3-DSBFs. The algorithm has produced 152
functions with the nonlinearity 241, and 36 many 3-DSBFs having nonlinearity 242.

Additionally, we have applied the search strategy to 9-variable 3-RSBFs
(the size of the search space is now $2^{176}$ as can be seen from Table 2), for
which we initiate the search algorithm with a 9-variable 3-DSBF having
nonlinearity 242. Then we have obtained some 9-variable 3-RSBFs (which are not in 3-DSBFs) having
nonlinearity 242, absolute indicator 56, and algebraic degree 7. The following
is the truth table of such a function:

\begin{center}
\ttfamily{3740B6A118A1E19642A85E2B7E2F3C3CB65FA0D95EC9DB1EA92BDB3666185AE0\\
087F5FE6E0757106A12FC918754C40E8A1BCCB7A714032A8961456E066E8A801\\}
\end{center}

It is clear that using one of the above 9-variable functions (say $f$) and a 2-variable bent function (say $g$), the 11-variable function $g(y_0, y_1) \oplus f(x_0, \ldots, x_8)$
with highest -till date- nonlinearity of $2^{11-1} - 2^{\frac{11-1}{2}} + 4 = 996$, can be
obtained. Similarly $h(y_0, y_1, y_2, y_3) \oplus f (x_0, \ldots, x_8)$ is the most nonlinear 13-variable function known to date, with nonlinearity $2^{13-1} - 2^{\frac{13-1}{2}} + 8 = 4040$ where $h$ is a 4-variable bent function and $f$ is one of the above 9-variable
functions with nonlinearity 242. We think this is a significant improvement on
the results of~\cite{kavut}. However, since the nonlinearity ($2^{n-1} - 2^{\frac{n-1}{2}} + 2\times2^{\frac{n-9}{2}}$), which can be obtained by bent concatenation of 9-variable functions with nonlinearity 242 is less than the nonlinearity ($2^{n-1} - 2^{\frac{n-1}{2}} + 20\times2^{\frac{n-15}{2}}$) given in~\cite{PW83} for odd $n\geq 15$, this result is significant only for odd $13\geq n\geq 9$.

\subsection{11 and 13-variable DSBFs}

In~\cite{MSD07}, the class of Dihedral Symmetric Boolean Functions (DSBFs), a subset of the RSBF class, which is invariant under the action of the dihedral group, is introduced. It has been shown that some of the 9-variable RSBFs having nonlinearity 241 also belong to this subset, demonstrating the richness of DSBFs in terms of high nonlinearity. Motivated by this, we have carried out a systematic search in the DSBF class for $15 >$ odd $n > 9$, and found Boolean functions having nonlinearity $> 2^{n-1}-2^\frac{n-1}{2}$. More specifically, for 11-variable DSBFs, we have attained an 11-variable DSBF with nonlinearity 994 within the space of size $2^{126}$. For 13-variable DSBFs, in order to reduce the search space ($2^{380}$), we have applied some additional permutations on input vectors, and obtained a subset of size $2^{74}$, in which we have found several 13-variable DSBFs with nonlinearity 4036. Consequently, our trials confirm that the DSBF class contains highly nonlinear Boolean functions and it is a rich subset of the RSBF class for $n = 11, 13$, as well. We should also mention that, this is the first demonstration of Boolean functions on odd number of input variables $9 < n < 15$ having nonlinearity greater than the bent concatenation bound, which are not obtained by the bent concatenation of 9-variable Boolean functions with nonlinearity $> 240$.

For the 11-variable DSBF case for which the size of search space is $2^{126}$, we have carried out 8000 runs of the search algorithm, and found the following 11-variable DSBF having nonlinearity 994, absolute indicator value 200, and algebraic degree 9: 

\begin{center}
\ttfamily{68C1F052AA14260999DD0365487844C6C397A7B6114A787724957BC46471F12D\\
F05F873ECC6E8A29034265887BD17A2A483583367B8FF1312C347E12FA1708F3\\
AA1433AF952B5BE9B5F02CA891985C92114A640C2D6380C57B9BB3027E991D8D\\
34C45B66D00E5B7D6ACF80EEAB021A430CE54E707AAD520DAB9D472F4081FF1F\\
89CC06215F1B8CAA973658CF27DAADD3CF36AA0118B0DDC08716D3D526E4C70D\\
065371D97C2054E458A2390BD550E5736ADAC2DF8B0A10492BACC3C317B381F7\\
1A21F52076CB3C3DB60144F836DF2AB32DDDE0EAC051FCBD8C8F10491299751F\\
41F0E96761AC6F053F888DE7234945F79C9B92B3703B19BF6545C557BBBF57FF
}
\end{center}

\noindent A typical run of the search algorithm takes 1 minute and 16 seconds using the same computer system. 

For the 13-variable DSBF case, since its search space is huge ($2^{380}$), before starting the search we apply the following permutation in addition to the permutations of dihedral group on input vectors $$\pi(x_0, x_1, \ldots, x_{12}) = (x_0, x_2, x_4, x_6, x_8, x_{10}, x_{12}, x_1, x_3, x_5, x_7, x_9, x_{11})$$

\noindent such that for each input $(x_0, \ldots, x_{12}) \in  \{0, 1\}^{13}$, $$f({\rho^1}_n(x_0, \ldots,x_{12})) = f(\tau_n(x_0, \ldots, x_{12})) = f(\pi(x_0, \ldots, x_{12})) = f(x_0, \ldots, x_{12}),$$ 

\noindent and the search space of 13-variable DSBFs is reduced from $2^{380}$ to $2^{74}$. Note that this permutation constitutes a subset of 13-variable DSBFs for which, using similar combinatorial methods as in~\cite{KMSY}, it may be possible to carry out an exhaustive search to enumerate 13-variable DSBFs with nonlinearity $\geq 4036$, with a reasonable amount of computational power. We have carried out 500 runs of the search algorithm, and found two 13-variable DSBFs having nonlinearity 4036 in this subset. One of them with nonlinearity 4036, absolute indicator value 208, and algebraic degree 10 is given below: 

\begin{center}
\ttfamily{177E7EF97EFCFF937FF8EBA0FAFBC71A7EFBEAD0EC8B8815EA99FADEA12A568D\\
7EE8EA8BF889B215FDB1848F80950677EDC883D3AE9DB2ED9D031888277CD4F7\\
7FEDE881ECDC948AFF90D0968B0C0676EFE3CE028524D4FAC114C666116C2A6B\\
F8E2A195815AF71A89FCD3A29B48BDE3C7F6155F139090904C2B2AA1F321AA3F\\
7EEEF8B2E881D107ECA1E3B5C665D088EAAE9354A710C37C81CB04E4156C3A28\\
ECAFEC0AB5ED504C85361D75B325AA88F4560730A4386C7C13537CF04CCD299B\\
FA85B81D9C129772D143368CFE2A43C88096AEB4B35E8809D3DE64959BE7A90E\\
A12BBF7D077227FA034FC601D340931535A159CF4C88CC17BB0B4D13C8990BAE\\
7EFCE8ACEAC18B0DE881C517E253407FF8F4C917EC5E9F32A12C3826F700C081\\
F889C9F8934B2770DC7B5710F44F2EF09146A5CA1530BD3107663CA14BCD0C81\\
F9A18DABB9F411C88A26ECF6364474B484321F7D47E33B779B5E58679CCD85D5\\
FA35626C042E4E419C244A902CF12EF5420E660B6EA0EE0570A0A4B64D86979E\\
FFCDD1728BC516A786F10348976A7F09B212350A0A78D5F1BFA85CD8350BA194\\
C015D72899F98F208E1B73E9C0950093B24AE7B96C65933782CAFC7BCCD715BC\\
9D0208CA8AAA7FE2147A2E49192BFBCD145A74FAF0790003E75B7451930B1736\\
1E76880372D3A1AB70F18590E1F5177A8E8B449F61B2075AF08597D6519ECCE9\\
7EEDEEF5FC90CDB4ED8CA10785CE10E3E881C147F523072FB81D274B71403EEF\\
FB81AE60F4C6122EB9A032EC96AA5A09C8171CF41ED05879BE7F5444A101D107\\
EAC481D3A197ABC0825E349E192A7A40A3A07BCA367F5300EE3424AE5DECAB15\\
C347647CD962E09806265E01DEA75B12117F3C3C1BA1D85771DBA0A751B58512\\
EE82891381B78D8FCE97AE741242F081C19C593CEDF4EB2C5A3874753A619B21\\
D1315E5802AE6AE6247FB95F5ECB6B3FC78A22A962842D2F82A1A0A3C026B276\\
FBDC1B632D1878F144700CFD70EC711687F05D3025D9870548B5AB5708EDAF76\\
700800F86C2951DB6CECCD01BDED51766E01CD11CD349F7C21A7943CC37ED6A9\\
EBAFE4B6B2133A49819BF1334739D86BD128EA42504A60C1876E6CCD6FEA11D3\\
8A1D17180E6650C810D86FD0A622FA179BAED88422E1E3D45F6651CFDD03D734\\
A0151733A72E48C196D6BE92C4EF4951D5EC028A2F5FE997B00182330101824E\\
DE5934CDAC2FCAD63CF43D33871F0B3EC00CA1DDAAB17BDEA1B5B67E13669EE1\\
93A6454915D5B5C9D088C8893AABB85D07742AC84CBC20C752C2099EFADBF1F6\\
077532896E64AB9DFA003F974105110AED6B238E6E753716821F05DE176E5A69\\
52B97F79C081414F3E08A60BC816CDDE3F41AA17C0779350B912AF76073E7AC9\\
C5FD819B602186BE79078F5C543E36C9BF158126D33EE6697712D6A9F4E1E997
}
\end{center}

\noindent In this case, a typical run takes one minute using the same computer system. Since these results confirm that the DSBF class contains highly nonlinear Boolean functions on 11 and 13-variables as well, it would be an interesting and open problem to attain some rich subsets achieving higher nonlinearity in the DSBF class. 

\section{Permutations on Input Vectors of 9-variable\\ Boolean Functions}

\noindent As it is deduced from the discussion in the preceding section, RSBFs are invariant under a special type of permutation. To search for better cryptographic characteristics, we consider the possibility of other classes of Boolean functions that are invariant under some permutations. Since linearly equivalent functions have the same nonlinearity, it makes sense to classify all $n!$ permutations up to the linear equivalence of Boolean functions that are invariant under them. The classification is based on the following proposition, which is easy to prove.

\begin{proposition} 
Let f and g be Boolean functions which are invariant under arbitrary permutations $\pi_f$ and $\pi_g$ respectively. Then, f and g are said to be linearly equivalent if there exists a bijective linear mapping $L: \{0, 1\}^n \rightarrow  \{0, 1\}$ such that $\pi_f = (L^{-1} \circ  \pi_g \circ  L)$. 
\end{proposition}

\noindent \textbf{Proof}. Suppose $f(x)=g(L(x))$, i.e., $f = g \circ L$. Then, it holds that $$f=g \circ L=g \circ \pi_g \circ L = f \circ L^{-1} \circ \pi_g \circ L=f \circ \pi_f.$$

Thus, we classify all possible permutations up to the equivalence $$\pi_f \sim \pi_g \Leftrightarrow  \exists L \mbox{ such that }  \pi_f = (L^{-1} \circ \pi_g \circ L).$$

The classification can be accomplished through a computer program by exploiting the Jordan Normal Form for matrices. Specifically for 9-variable Boolean functions, all permutations of the identity matrix ($362,880$ many) yield that there are only 30 permutations (see Table 3), which are different up to the equivalence defined above. Then, we apply the search algorithm for each class using its representative permutation and determine the corresponding nonlinearity given in the last column of Table 3. Our results show the existence of permutations having similar cryptographic characteristics with $k$-RSBFs and $k$-DSBFs.   

From Table 3, it is seen that we have attained several 9-variable Boolean functions with nonlinearity 242, which we initially found in 3-DSBFs and 3-RSBFs, in the classes with sizes $2^{100}$, $2^{104}$, $2^{140}$. In the following, we present 9-variable Boolean functions having nonlinearity 242 and different autocorrelation spectra from those of the functions found in 3-DSBFs and 3-RSBFs.

\normalsize{}We have applied 100 runs of the search algorithm~\cite{kavut} to the space of size $2^{104}$ and found two 9-variable Boolean functions with nonlinearity 242, absolute indicator value 48, and algebraic degree 7. A typical run takes the same amount of time as for the case of 3-DSBFs (since the sizes of both spaces are the same). One of these functions is given below:

\begin{center}
\ttfamily{7B8F94BAD364DAC9931906F9465FF33E921E13D7552DAFD684757B662FDA3C68\\
FA8D94B3C3659B5FCC46FD1518050F97A1E02039AAF74337134F30AB5B41D9DE}
\end{center}

\noindent which is invariant under the representative permutation $$\pi(x_0, x_1, \ldots, x_8) = (x_0, x_2, x_1, x_4, x_5, x_6, x_7, x_8, x_3).$$
\newpage

\noindent \small{{\textbf{Table 3}. Classification of all possible $362,880$ many permutations for 9-variable Boolean functions, and the best achieved nonlinearity result for each class.\begin{center} 
\begin{tabular}{| c | c | c | c | c |}

  \hline
                         &                      & \small{Maximum}        & \small{Total}      &                        \\
                         &                      & \small{number of}      & \small{number of}  & \small{Best achieved} \\
  \small{Representative} & \small{Number of}    & \small{input vectors}  & \small{distinct}   & \small{nonlinearity}  \\ 
  \small{permutation}    & \small{permutations} & \small{in an orbit}    & \small{orbits}     & \small{result}        \\ \hline
  (0,1,2,3,4,5,6,7,8) &   &   &     &     \\ 
  (\textit{identity}) & 1 & 1 & 512 & 239 \\ \hline
  (5,7,4,8,2,0,6,1,3) & 945					& 2 & 272 & $240^6$ \\ \hline
	(3,1,7,0,5,4,6,2,8) & 1260  &	2  &	288 &	240 \\ \hline
	(7,1,2,3,5,4,6,0,8) & 378   &	2  &	320 &	240 \\ \hline
	(0,8,2,3,4,5,6,7,1) & 36    &	2  &	384 &	240 \\ \hline
	\textbf{(4,6,7,2,8,1,5,3,0)} & \textbf{2240}  &	\textbf{3}  &	\textbf{176} &	$\textbf{242}^{3,4}$ \\ \hline
	(5,1,2,4,7,8,6,3,0) & 3360  &	3  &	192 &	240 \\ \hline
	(0,1,2,8,4,5,6,3,7) & 168   &	3  &	256 &	240 \\ \hline
	\textbf{(1,4,7,5,6,2,0,3,8)} & \textbf{11340} &	\textbf{4}  &	\textbf{140} & $\textbf{242}^2$ \\ \hline
	(4,7,5,6,0,1,3,2,8) & 11340 &	4  &	168 &	240 \\ \hline
	(0,2,1,7,4,3,5,6,8) & 7560  &	4  &	176 &	240 \\ \hline
	(0,8,2,3,4,1,6,5,7) & 756   &	4  &	192 &	240 \\ \hline
	(0,7,2,5,4,1,3,6,8) & 3024  &	5  &	128 &	239 \\ \hline
	\textbf{(8,7,3,0,1,6,2,4,5)} & \textbf{20160} &	\textbf{6}  &	\textbf{100} & $\textbf{242}^1$ \\ \hline
	\textbf{(0,2,1,4,5,6,7,8,3)} & \textbf{30240} &	\textbf{6}  &	\textbf{104} &	\textbf{242} \\ \hline
	(7,1,0,3,4,2,8,6,5) & 10080 &	6  &	112 &	240 \\ \hline
	(7,4,0,5,1,8,6,2,3) & 10080 &	6  &	144 &	240 \\ \hline
	(8,4,3,2,1,7,5,6,0) & 2520  &	6  &	144 &	240 \\ \hline
	(0,6,2,7,8,1,5,3,4) & 7560  &	6  &	160 &	240 \\ \hline
	(8,1,3,2,4,5,0,7,6) & 2520  &	6  &	192 &	240 \\ \hline
	(2,0,6,1,4,5,7,8,3) & 25920 &	7  &	80  & 240 \\ \hline
	(0,3,5,8,1,4,7,2,6) & 45360 &	8  &	72  &	240 \\ \hline
	\textbf{(1,6,7,4,8,2,5,3,0)} & \textbf{40320} &	\textbf{9}  &	\textbf{60}  & $\textbf{241}^5$ \\ \hline
	(5,8,6,7,2,0,1,3,4) & 9072  &	10 &	80  & 240 \\ \hline
	(1,3,8,7,4,0,6,5,2) & 18144 &	10 &	96	& 240 \\ \hline
	(3,5,7,1,6,0,8,2,4) & 15120 &	12 &	88	& 240 \\ \hline
	(0,2,7,8,4,6,3,1,5) & 15120 &	12 &	96	& 240 \\ \hline
	(4,5,7,1,0,8,3,6,2) & 25920 &	14 &	60	& 240 \\ \hline
	(6,5,1,4,7,2,3,0,8) & 24192 &	15 &	64	& 238 \\ \hline
	(4,8,1,2,6,7,5,0,3) & 18144 &	20 &	48	& 240 \\ 
  
  \hline
\end{tabular}	
\end{center}	
\footnotesize{$^1$ Nonlinearity result of 242 is attained in the subset of size $2^{74}$ in the set of size $2^{100}$.}\\
\footnotesize{$^2$ Nonlinearity result of 242 is attained in the subset of size $2^{86}$ in the set of size $2^{140}$.}\\
\footnotesize{$^3$ Nonlinearity result of 242 is attained in the subset of size $2^{104}$ in the set of size $2^{176}$.}\\
\footnotesize{$^4$ The class contains the permutation corresponding to 3-RSBFs.}\\ 
\footnotesize{$^5$ The class contains the permutation corresponding to RSBFs.}\\ 
\footnotesize{$^6$ The class contains the permutation corresponding to 9-DSBFs.}\\

\normalsize{}Then, in order to reduce the search space, we have considered some subclasses. For this purpose, we have applied the reflection operator, which is defined as  $\tau_n(x_0, x_1, \ldots , x_8) = (x_8, \ldots , x_1, x_0)$ for 9-variable Boolean functions, in addition to the representative permutation. As a result of this method, we have identified a subset of size $2^{74}$ in the set of size $2^{100}$. In this subset, we have attained several 9-variable Boolean functions with nonlinearity 242, absolute indicators 40 and 64, and algebraic degree 7. One of them having absolute indicator 64 is provided below:	

\begin{center}
\ttfamily{0331786B34D878855663A2E961F1CB4F779EBBF6881ABB24AC033E6C2B32E049\\
3D0891DB1888EA5E6F910310311532FC68D5F2A4B5BE6445E41F64299F0CC99A}
\end{center}

\noindent which is invariant under the permutations of the reflection operator  $\tau_n$ and the representative permutation $$\pi(x_0, x_1, \ldots, x_8) = (x_8, x_7, x_3, x_0, x_1, x_6, x_2, x_4, x_5).$$

\noindent For this case, we have carried out 100 runs of the search algorithm resulting in 9 many Boolean functions with nonlinearity 242 such that seven of them with absolute indicator value of 64, and the remaining with that of 40. A typical run takes 1 minute and 4 seconds using the same computer system. 

\section{Conclusions}

\noindent By suitably generalizing the class of RSBFs, we have introduced $k$-RSBFs, as functions which satisfy $f(\alpha^{2^k}) = f(\alpha)$, where the nonzero positive integer $k$ divides $n$, and $\alpha \in GF(2^n)$. We have also defined the class of $k$-DSBFs as a subset of $k$-RSBFs imposing the condition of invariance under the action of dihedral group. Using the steepest-descent like iterative algorithm in~\cite{kavut, KMY06} for a search in the generalized 3-DSBF and 3-RSBF classes, we have attained 9-variable 3-RSBFs and 3-DSBFs with nonlinearity 242. This result shows the existence of $n$-variable Boolean functions having nonlinearity $2^{n-1}-2^\frac{n-1}{2}+2\times 2^\frac{n-9}{2}$ for $n = 9, 11, 13$. In~\cite{BR92}, it is conjectured that the covering radius~\cite{FJ77, PW83} of $R$(1, $n$) is even. Our nonlinearity result for $n = 9$ shows that the covering radius is at least 242 and it is an interesting open question to settle it. The upper bounds presented in~\cite{TOR78, HO97} for the covering radius of $R$(1, 9) is 244.

Further, we have classified all possible permutations up to the linear equivalence of Boolean functions that are invariant under them. Specifically for $n = 9$, there are 30 such classes. Exploiting the same search algorithm~\cite{kavut}, we have attained 9-variable Boolean functions having nonlinearity 242 in the class of size $2^{104}$. Then, we have considered some subclasses by adding permutation of the reflection operator $\tau_n$ to the representative permutation and found many functions with nonlinearity 242 shown in Table 3. As an example, we have identified a subset of size as small as $2^{74}$, in the set of size $2^{100}$, having 9-variable Boolean functions with nonlinearity 242. Considering the combinatorial search techniques in~\cite{KMSY}, we note that it may be possible to exhaustively search the subset of size $2^{74}$ for the enumeration of 9-variable Boolean functions having nonlinearity $\geq 242$, with a reasonable amount of computational power. Moreover, we have obtained an 11-variable DSBF having nonlinearity 994 and several 13-variable DSBFs having nonlinearity 4036, which confirm the richness of DSBFs~\cite{MSD07} in terms of high nonlinearity for $n = 11$ and 13. 

We think that the results that we present contain significant information on the existence of maximum nonlinearity-Boolean functions with odd number of input variables, within the classes that are invariant under some permutations.\\

\noindent {\bf Acknowledgment.} 
We acknowledge Dr. Gregor Leander for his discussion on linear equivalence of Boolean functions that are invariant under random permutations, which has helped us find one missing class in Table 3.

\end{document}